\documentclass[preprint,pra,showpacs,nofootinbib]{revtex4}
\usepackage{mathrsfs}
\usepackage{graphicx}
\usepackage{epsfig}
\usepackage{dcolumn}% Align table columns on decimal point
\usepackage{bm}% bold math
\usepackage{amsmath,amssymb,amsthm}
\usepackage[colorlinks=true,linkcolor=blue]{hyperref}
\usepackage{subfigure}
\usepackage{booktabs}
\usepackage[mathscr]{euscript}
\usepackage[normalem]{ulem}
\usepackage{xcolor}

\newcommand\redsout{\bgroup\markoverwith{\textcolor{red}{\rule[0.6ex]{6pt}{0.6pt}}}\ULon}

\newtheorem{prop}{Proposition}
\newcommand{\bea}{\begin{eqnarray}}
\newcommand{\eea}{\end{eqnarray}}
\newcommand{\beq}{\begin{equation}}
\newcommand{\eeq}{\end{equation}}

\def\/{\over}

\begin{document}

\title{Duality of positive and negative integrable hierarchies via relativistically invariant fields}
\author{S. Y. Lou$^{1}$, X. B. Hu$^{2,3}$ and Q. P. Liu$^{4}$
\thanks{Corresponding author:lousenyue@nbu.edu.cn. Data Availability Statement: The data that support the findings of this study are available from the corresponding author upon reasonable request.}}
\affiliation{
$^1$\mbox {\scriptsize School of Physical Science and Technology, Ningbo University, Ningbo, 315211, China }
$^{2}$\mbox {\scriptsize  LSEC, ICMSEC,  Academy of Mathematics and Systems Science, Chinese Academy of Sciences, Beijing 100190, China }
$^{3}$\mbox {\scriptsize School of Mathematical Sciences, University of Chinese Academy of Sciences, Beijing 100049, China}
$^{4}$\mbox {\scriptsize Department of Mathematics, China University of Mining and Technology, Beijing 100083, China }}

\begin{abstract}
  It is shown that the relativistic invariance plays a key role in the study of integrable systems. Using the relativistically invariant sine-Gordon equation, the Tzitzeica equation, the Toda fields and the second heavenly equation as dual relations, some continuous and discrete integrable positive hierarchies such as the potential modified Korteweg-de Vries hierarchy, the potential Fordy-Gibbons hierarchies, the potential dispersionless Kadomtsev-Petviashvili-like (dKPL) hierarchy, the differential-difference dKPL hierarchy and the second heavenly hierarchies are converted to the integrable negative hierarchies including the sG hierarchy and the Tzitzeica hierarchy, the two-dimensional dispersionless Toda hierarchy, the two-dimensional  Toda hierarchies and negative heavenly hierarchy. In (1+1)-dimensional cases the positive/negative hierarchy dualities are guaranteed by the dualities between the recursion operators and their inverses. In (2+1)-dimensional cases, the positive/negative hierarchy dualities are explicitly shown by using the formal series symmetry approach, the mastersymmetry method and the relativistic invariance of the duality relations. For the 4-dimensional heavenly system, the duality problem is studied firstly by formal series symmetry approach. Two elegant commuting recursion operators of the heavenly equation appear naturally from the formal series symmetry approach so that the duality problem can also be studied by means of the recursion operators.\\
\leftline{}
%02.30.Ik (Integrable systems), 11.10.Lm (Nonlinear or nonlocal theories and models), 11.30.-j(Symmetries and conservation laws), 03.50.-z (classical field theories), 03.50.-z (other special classical field theories), 03.30.+p (special relativity) 05.45.Yv Solitons, 05.45.-a nonlinear dynamics, 11.30.Na Nonlinear and dynamical symmetries (spectrum-generating symmetries)
\end{abstract}

\maketitle

\section{Introduction}
The study on dualities such as the wave-particle duality  in quantum physics \cite{WaveP}, the electromagnetic duality between electric and magnetic fields \cite{DianCi}, AdS/CFT duality (duality between the Anti-de Sitter gravity and the conformal field theory) \cite{AdS}, the particle-vortex duality in condensed matter physics \cite{Ann}, the boson/fermion dualities  in particle physics \cite{PRX,GaoXN} and quantum/classical duality between quantum and classical integrable fields \cite{QC} is an important topic in Physics. In this paper, we study the positive/negative hierarchy dualities, the dualities between integrable positive hierarchies and negative hierarchies by means of the relativistically invariant fields, such as the sine-Gordon (sG) field, the Tzitzeica field, the Toda fields and the heavenly equations.

In (1+1)-dimensional cases, the positive/negative hierarchy dualities are equivalent to the local/nonlocal symmetry dualities for the related integrable systems.
The investigation of symmetries plays a fundamental role in natural science. It is known that the standard model including all the known elementary particles constituted our universe is based on the local SU(3)$\times$SU(2)$\times$U(1) gauge symmetry \cite{Gauge1,Gauge2,LouNi}. All the predictions resulted from the standard model have been confirmed  whence the Higgs boson is found \cite{Higgs}. In nonlinear science, there are numerous models to describe real natural phenomena. While there is no unified method to solve nonlinear systems, the symmetry approach has been very effective. Symmetries have numerous applications \cite{Clarkson},  including to build new solutions from known ones \cite{YQLi,TXY}, to do dimensional reductions of nonlinear partial differential equations \cite{Clarkson1,Lou}, to get new integrable systems \cite{Cao,Cheng,Strampp,LouHu} and even to construct all solutions for certain nonlinear systems \cite{LouYao}.

For a (1+1)-dimensional integrable system, there are infinitely many local and nonlocal symmetries. To produce a hierarchy of infinitely many local commuting symmetries, one can apply  a recursion operator \cite{Olver}, $\Phi$, to a seed symmetry which usually lies in a kernel of the inverse recursion operator $\Phi^{-1}$. To find a set of infinitely many nonlocal symmetries, some different approaches are available. The  simplest way is to apply the inverse of recursion operator, $\Phi^{-1}$, to a seed symmetry which belongs to  a kernel of the recursion operator $\Phi$ \cite{LouPLB}. In addition to the kernels of the recursion operators, there are many other ways to find seed nonlocal symmetries, say, the squared eigenfunction symmetries \cite{eigenF} related to Lax pairs, the residual symmetries of the truncated Painlev\'e expansions \cite{GaoXN}, the infinitesimal Darboux transformations \cite{DT1,LouHu,DT3}, the infinitesimal B\"acklund transformations \cite{BT1NH2,BT2} and the infinitesimal conformal transformations  of the Schwarzian forms \cite{CT}.

For a (2+1)-dimensional integrable model, it is much more difficult to find symmetries because the nonexistence of recursion operator except for C-integrable models (the models which can be directly linearized) and breaking soliton models (whose recursion operators are of one-dimensional). To find the symmetries related to (2+1)-dimensional positive hierarchies, the mastersymmetry method \cite{Fuch} and the formal series symmetry approach (FSSA) \cite{FSSA} are two effective methods. To find (2+1)-dimensional negative integrable hierarchies, one may apply the formal spectral parameter expansion method \cite{NH1,BT1NH2,LouHu} to suitable seed nonlocal symmetries like the squared eigenfunction symmetries, residual symmetries and infinitesimal Darboux/B\"acklund transformations.

To describe the colourful real nonlinear natural world, there are various idealized models including the Korteweg-de Vries (KdV) equation \cite{KdV}, the modified KdV (MKdV) equation \cite{MKdV}, the potential KdV (PKdV) equation, the potential MKdV (PMKdV) equation, the sine-Gordon (sG) equation \cite{sG}, the Sawada-Kotera (SK) equation \cite{SK}, the Kaup-Kupershmidt (KK) equation \cite{KK}, the Fordy-Gibbons (FG) equation \cite{FG}, the potential FG (PFG) equation, the Tzitzeica equation \cite{TE,TE1,DDToda1}, the nonlinear Schr\"odinger (NLS) equation \cite{NLS} and so on. It is interesting to point out that some of these physically relevant models are linked each other. Indeed, the MKdV equation is related to the KdV equation by a Miura transformation. The PMKdV equation is a potential form of the MKdV equation. The SK equation and the KK equation are linked to a common modified equation, the FG equation, via different Miura transformations \cite{FG}. The KdV equation, the MKdV equation, the NLS equation and the sG equation are all the special reductions of the Ablowitz-Kaup-Newell-Segur hierarchy \cite{AKNS}. The PMKdV equation and the sG equation possess a same recursion operator and some sets of common infinitely many conserved invariants and symmetries. The PFG equation and the Tzitzeica equation share a common recursion operator and some sets of invariants and symmetries.
The sG equation and the Tzitzeica equation are all relativistically invariant, invariant under the Lorenz transformation in the experimental coordinate system or invariant under the space-time exchange in the light cone coordinate. In this paper, we study another type of connection among these integrable systems by means of the relativistically invariant equations like the sG, Tzitzeica, Toda and heavenly equations.

The paper is organized as follows. In section II, with the help of the relativistic invariance of the sG equation, we study the local/nonlocal symmetry duality or equivalently the positive/negative hierarchy duality related to the PMKdV equation and the sG equation, i.e., the duality of the PMKdV hierarchy and the sG hierarchy. The duality of the PFG hierarchies and the Tzitzeica hierarchies are established in section III thanks to the relativistic invariance of the Tzitzeica equation. In section IV, after reviewing the FSSA and then applying the method to a (2+1)-dimensional potential dKPL model, the positive/negative hierarchy duality is established by using the two-dimensional dispersionless Toda (2ddT) equation. In section V, the duality method is treated alternatively. We directly use the relativistically invariant differential-difference Toda equation to find a higher order Toda system and its related dual model. In section VI, the hierarchy duality problem is studied for a four dimensional integrable model (the second heavenly equation) by means of the FSSA. Two commute elegant recursion operators of the second heavenly equation are reobtained from the FSSA. Then, the positive/negative heavenly hierarchy duality is established with the helps of the recursion operators and the relativistic invariance of the heavenly equation. The last section is a short summary and some discussions.

\section{Duality of the PMKdV hierarchy and the sine-Gordon hierarchy}

The sG equation (in the light cone coordinate)
\begin{equation}
v_{x\tau}=\sin (v) \label{sG}
\end{equation}
 is one of the most physically relevant  field equations \cite{sG,sG1,sG11,sG2,sG21,sG3,sG31}. It is important not only in quantum and classical field theories but also in almost all the physical branches and even in other natural scientific fields. For instance, the sG model is equivalent to the massive Thirring model \cite{Thirring1,T1}, the two-dimensional Coulomb gas \cite{sG2,sG4}, the continuous limit of lattice $x$-$y$-$z$ spin-half model \cite{xyz}, and the massive $O$(2) nonlinear $\sigma$ model \cite{sG2}.

It is known that the sG model \eqref{sG} and the PMKdV equation
\begin{equation}
v_{t}=v_{xxx}+\frac12v_x^3 \label{PMKdV}
\end{equation}
share a common recursion operator
\begin{equation}
\Phi=\partial_x^2+v_x^2-v_x\partial_x^{-1}v_{xx}, \label{RO_PMKdV}
\end{equation}
where $\partial^{-1}_x$ is formally defined by $\partial_x^{-1}\partial_x=\partial_x\partial_x^{-1}=1$. When the operator, $\partial^{-1}_x$, is restricted to act on the functions with vanishing boundary at $x=-\infty$, one can simply define it by $\partial_x^{-1}=\int_{-\infty}^x \mbox{\rm d}y$ \cite{SasakiB}.

Usually, the inverse recursion operator of $\Phi$ is written as \cite{SasakiB,PLB93,JMP94}
\begin{equation}
\Phi^{-1}=\partial_{\sin}^{-2}+\partial_{\cos}^{-2}=\frac12\left(\partial_{\mbox{\rm e}^{iv}}^{-1}\partial_{\mbox{\rm e}^{-iv}}^{-1}+\partial_{\mbox{\rm e}^{-iv}}^{-1}\partial_{\mbox{\rm e}^{iv}}^{-1}\right), \label{RO_sG}
\end{equation}
where the operators $\partial_{\sin}^{-1}$, $\partial_{\cos}^{-1}$, $ \partial_{\mbox{\rm e}^{iv}}^{-1}$ and $\partial_{\mbox{\rm e}^{-iv}}^{-1}$ are defined by
\begin{equation}
\partial_{\sin}^{-1} f\equiv \partial_x^{-1}[\sin(v)f],\ \partial_{\cos}^{-1} f\equiv \partial_x^{-1}[\cos(v)f],\ \partial_{\mbox{\rm e}^{iv}}^{-1}f\equiv\partial_x^{-1}[\mbox{\rm e}^{iv}f],\ \partial_{\mbox{\rm e}^{-iv}}^{-1}f\equiv\partial_x^{-1}[\mbox{\rm e}^{-iv}f] \label{partial1}
\end{equation}
for an  arbitrary function $f$. Thus, the PMKdV hierarchy (positive hierarchy) and the sG hierarchy (negative hierarchy) can be written as
\begin{equation}
v_{t_{2n+1}}=\Phi^nv_x=(\partial_x^2+v_x^2-v_x\partial_x^{-1}v_{xx})^nv_x\equiv K_{2n+1},\ n=0,\ 1,\ 2,\ \ldots, \label{PH}
\end{equation}
and
\begin{equation}
v_{\tau_{2n+1}}=\Phi^{-n}\partial_x^{-1} \sin(v)=(\partial_{\sin}^{-2}+\partial_{\cos}^{-2})^{n}\partial_x^{-1} \sin(v)\equiv K_{-2n-1},\ n=0,\ 1,\ 2,\ \ldots, \label{NH}
\end{equation}
respectively. It is known \cite{SasakiB} that $K_{2n+1}$ defined in \eqref{PH} are local symmetries and $K_{-2n-1}$ defined in \eqref{NH} are nonlocal symmetries of the PMKdV equation \eqref{PMKdV} for all $n=0,\ 1,\ 2,\ \ldots$.

First few flow equations of \eqref{PH} and \eqref{NH} are listed as follows
\begin{eqnarray}
&&v_{t_1}=v_x,\label{vt1}\\
&&v_{\tau_1}\equiv v_{\tau}=\partial_x^{-1}\sin (v),\label{vT1}
\end{eqnarray}
\begin{eqnarray}
&&v_{t_3}\equiv v_{t}=v_{xxx}+\frac12v_x^3,\label{vt3}\\
&&v_{\tau_3}=\frac16v_{\tau}^3+\partial_{\cos}^{-2}v_{\tau}, \
v_{\tau}=\partial_x^{-1} \sin (v)\label{vT3}
\end{eqnarray}
\begin{eqnarray}
&& v_{t_5}=v_{xxxxx}+\frac52v_x^2v_{xxx}+\frac52v_xv_{xx}^2+\frac38v_x^5,\label{vt5}\\
&&v_{\tau_5}=\frac1{120}v_{\tau}^5+\frac16\partial_{\cos}^{-2}v_{\tau}^3
+\left(\partial_{\sin}^{-2}+\partial_{\cos}^{-2}\right)\partial_{\cos}^{-2}v_{\tau},\label{vT5}
\end{eqnarray}
\begin{eqnarray}
&& v_{t_7}=v_{xxxxxxx}+\frac72v_x^2v_{xxxxx}+14v_xv_{xx}v_{xxxx}+\frac{21}2v_xv_{xxx}^2
+\frac{35}8v_x^4v_{xxx}\nonumber\\
&&\qquad\quad+\frac{35}4(2v_{xxx}+v_x^3)v_{xx}^2+\frac5{16}v_x^7,\label{vt7}\\
&&v_{\tau_7}=\frac1{5040}v_{\tau}^7+\frac1{120}\partial_{\cos}^{-2}v_{\tau}^5+\frac16\left(\partial_{\sin}^{-2}+\partial_{\cos}^{-2}\right)
\partial_{\cos}^{-2}v_{\tau}^3+\left(\partial_{\sin}^{-2}+\partial_{\cos}^{-2}\right)^2
\partial_{\cos}^{-2}v_{\tau}.\label{vT7}
\end{eqnarray}
Using the definition of the commutation relation $[F(v),\ G(v)]$ as 
$$[F(v),\ G(v)]=F'G-G'F=\lim_{\epsilon=0}\frac{\mbox{\rm d}}{\mbox{\rm d}\epsilon}[F(v+\epsilon G)-G(v+\epsilon F)], $$
it is not difficult to prove that all the local symmetries $\Phi^nv_x=K_{2n+1}$ and the nonlocal symmetries $\Phi^{-n}\partial_x^{-1} \sin(v)=K_{-2n-1}$ commute each other \cite{SasakiB,PLB93},
$$[K_{2n+1},\ K_{2m+1}]=[K_{2n+1},\ K_{-2m-1}]=[K_{-2n-1},\ K_{-2m-1}]=0,\ n,\ m=0,\ 1,\ 2,\ \ldots $$
which means everyone of $K_{2n+1}$, $K_{-2n-1}$ and their linear combinations are solutions of the symmetry equations of the PMKdV and the sG hierarchies. The symmetry equation of the PMKdV equation is defined as
\begin{eqnarray}
&&\sigma_{t}=\sigma_{xxx}+\frac32v_x^2\sigma_x,\label{st3}
\end{eqnarray}
which is the linearized equation of \eqref{vt3} and is obtained by substituting $v\rightarrow v+\epsilon \sigma $ with the infinitesimal parameter $\epsilon$ into the PMKdV equation \eqref{vt3}.

The  symmetry equation of the sG equation,
\begin{equation}
\sigma_{x\tau}=\cos (v)\sigma, \label{ssG}
\end{equation}
is the linearized form of Eq. \eqref{sG}.

Because the sG equation \eqref{sG} is relativistically invariant ($x,\ \tau$ exchange invariance) and $\Phi$ given by \eqref{RO_PMKdV} is a proper recursion operator for it, we immediately have the proposition 1.
\begin{prop}The operator
\begin{equation}
\Phi_1=\partial_{\tau}^2+v_{\tau}^2-v_{\tau}\partial_{\tau}^{-1}v_{\tau\tau} \label{RO_tau}
\end{equation}
is a recursion operator for the sG equation \eqref{sG}. \end{prop}

It is observed that  $\Phi_1$ is nothing but  the inverse of the recursion operator $\Phi$ as implied by the following proposition 2.
\begin{prop} If $\sigma$ is a symmetry of the sG equation \eqref{sG}, i.e., a solution of \eqref{ssG}, then
\begin{equation}
\Phi_1\Phi\sigma=\sigma =\Phi\Phi_1\sigma. \label{PhiPhi1}
\end{equation}\end{prop}
\bf Proof. \rm The second equality follows from the first one under the exchange $x\leftrightarrow \tau$, so it is enough to prove the first equation of \eqref{PhiPhi1}. Indeed, we have
\begin{eqnarray}
&&\Phi_1\Phi\sigma  =\big(\partial_x^2\partial_{\tau}^2+v_x^2\partial_{\tau}^2
+v_\tau^2\partial_{x}^2+
2 v_{x\tau}^2\big)\sigma+v_{x\tau\tau}(2v_x-\partial_x^{-1}v_{xx})\sigma
\nonumber\\
&&\qquad\qquad+2v_{x\tau}\left(2v_x\partial_{\tau}-\partial_x^{-1}\partial_{\tau}v_{xx}\right)\sigma
 +v_xv_{\tau}^2(v_x-\partial_x^{-1}v_{xx})\sigma
 \nonumber\\
&&\qquad\qquad
+v_{\tau}\partial_{\tau}^{-1}(v_xv_{\tau\tau}\partial_x^{-1}{v_{xx}}
-v_{\tau\tau}v_x^2-v_{\tau\tau}\partial_{x}^2)\sigma-v_x\partial_x^{-1}\partial_{\tau}^2 v_{xx}\sigma.
\label{proof}
\end{eqnarray}
Eliminating $v_{x\tau}$ and $\sigma_{x\tau}$ via the sG equation \eqref{sG} and its symmetry equation \eqref{ssG}, \eqref{proof} yields
\begin{eqnarray}
&&\Phi_1\Phi\sigma =[1+v_xv_{\tau}\cos(v)+v_x^2v_{\tau}^2+v_x\sin(v)\partial_{\tau}]\sigma
-v_xv_{\tau}^2\partial_x^{-1}v_{xx}\sigma
\nonumber\\
&&\qquad\qquad-v_x\partial_{\cos}^{-1}[\sin(v)+v_x\partial_{\tau}]\sigma
-v_\tau\cos(v)\partial_x^{-1}v_{xx}\sigma
 \nonumber\\
&&\qquad\qquad+v_{\tau}\left\{\partial_{\tau}^{-1}\left[v_{x}v_{\tau\tau}(\partial_{x}^{-1}v_{xx}-v_x)
-\sin(v)(\cos(v)
+v_xv_{\tau})\right]\right\}\sigma.
\label{proof1}
\end{eqnarray}
Performing some  integrations by parts and using the relations \eqref{sG} and \eqref{ssG}, it is straightforward to find that \eqref{proof1} is just the first equation of \eqref{PhiPhi1}, and the proposition 2 is proved.

Because $\Phi_1$ expressed in \eqref{RO_tau} is just the inverse of $\Phi$, the sG hierarchy \eqref{NH} (the negative PMKdV hierarchy) can be reformulated as
\begin{equation}
v_{\tau_{2n+1}}=(\partial_{\tau}^2+v_{\tau}^2-v_{\tau}\partial_{\tau}^{-1}v_{\tau\tau})^nv_{\tau},\ n=0,\ 1,\ 2,\ \ldots, \label{NH1}
\end{equation}
with $v_{\tau}=\partial_x^{-1}\sin(v)$.

Thus, we call the positive PMKdV hierarchy \eqref{PH} and the negative PMKdV hierarchy \eqref{NH} (sG hierarchy) are dual each other while the sG equation is termed as the duality relation of the hierarchies. In other words, the local symmetries $K_{2n+1}$ and the nonlocal symmetries $K_{-2n-1}$ are dual with the duality relation
\begin{equation}
v_x\leftrightarrow v_{\tau}[=\partial_x^{-1}\sin(v)].\label{dual}
 \end{equation}
 Alternatively, we can also say that the set of the nonlocal symmetries of the PMKdV equation can be localized with help of the duality relation \eqref{dual}. In fact, for the $n$th equation of the negative hierarchy  \eqref{NH} or \eqref{NH1} is local in $\{\tau, \tau_{2n+1}\}$ space-time ($\tau$ space and $\tau_{2n+1}$ time) and nonlocal in $\{x,\ \tau_{2n+1}\}$ space-time ($x$ space and $\tau_{2n+1}$ time). For the $n$th equation of the positive hierarchy \eqref{PH} is local in $\{x, t_{2n+1}\}$ space and nonlocal in $\{\tau,\ t_{2n+1}\}$ space while $x$ and $\tau$ are related by the sG equation.

Summarizing above results, we have the following conjecture.\\
\bf Dual conjecture of positive and negative hierarchies: \rm For an integrable system, there exists a possible dual relation such that a positive hierarchy can be changed to a negative hierarchy.

If the positive hierarchy (like the PMKdV hierarchy) is local and the negative hierarchy is nonlocal (like the sG hierarchy), then the duality conjecture indicates that the local symmetries and nonlocal symmetries may be dual each other via a possible dual relation.

\section{Duality of the PFG hierarchy and the Tzitzeica hierarchy}

To provide further support  for our conjecture, we consider another well known integrable system, the PFG equation \cite{FG},
\begin{equation}
v_t=v_{xxxxx}-\frac52v_{xx}v_{xxx}-\frac54v_xv_{xx}^2-\frac54v_x^2v_{xxx}+\frac1{16}v_x^5\equiv K_5.\label{PFG}
\end{equation}
 It is a potential form of the FG equation $(u=v_x)$ \cite{FG}
\begin{equation*}
u_t=\left(u_{xxxx}-\frac52u_{x}u_{xx}-\frac54uu_{x}^2-\frac54u^2u_{xx}+\frac1{16}u^5\right)_x,\label{FG}
\end{equation*}
which is a common modified system of the SK and KK equations via different Miura transformations. The SK and KK equations are two important physical models which can be used to describe all the physical fields where the KdV equation is not adequate and needs some higher order corrections \cite{JPC}.

From the results of the SK and KK equations, we know that the PFG equation possesses two sets of local symmetries, or equivalently, two positive hierarchies \cite{JMP94}
\begin{equation}
v_{t_{6n+1}}=\Psi_D^n v_x\equiv \Psi_D^n K_1, \ n=0,\ 1,\ 2,\ \ldots, \label{PFG1}
\end{equation}
and
\begin{equation}
v_{t_{6n+5}}=\Psi_D^n K_5,\ n=0,\ 1,\ 2,\ \ldots  \label{PFG2}
\end{equation}
with $K_5$ being defined in \eqref{PFG}, the recursion operator and the inverse recursion operator being given by
\begin{eqnarray}
\Psi_D&=&D_gg^{-1}D_{g^3}^{-1}D_g^5g^{-2}D_{g^2}^{-1}D_g^2,\ g\equiv \exp\left(\frac{v}2\right),   \label{Psi}\\
\Psi_D^{-1}&=&D_g^{-2}D_{g^2}g^{2}D_g^{-5}D_{g^3}g D_g^{-1}  \label{PsiN}\\
&=&\frac19\left(\partial_2^{-1}\partial_1^{-1}\partial_2^{-2}\partial_1^{-1}\partial_2^{-1} +2\partial_1^{-1} \partial_2^{-3}\partial_1^{-1}\partial_2^{-1} +2\partial_2^{-1}\partial_1^{-1}\partial_2^{-3}\partial_1^{-1}+4\partial_1^{-1}\partial_2^{-4} \partial_1^{-1}\right),\nonumber
\end{eqnarray}
where
\begin{equation}
 D_f\equiv f\partial_x,\ D_{f}^{-1}\equiv (D_f)^{-1}=\partial_x^{-1}f^{-1},\ \partial_1^{-1}\equiv \partial_x^{-1}\mbox{\rm e}^v,\ \partial_2^{-1}\equiv \partial_x^{-1}\mbox{\rm e}^{-\frac{v}2}.   \label{Df}
\end{equation}
It is mentioned that the seed symmetries $K_1$ and $K_5$ are only two nontrivial solutions of  $\Psi_D^{-1}f=0$.

 The PFG equation is known to possess a nonlocal symmetry $\sigma=\partial_x^{-1} \left(\mbox{\rm e}^v+\mbox{\rm e}^{-\frac{v}2}\right)$ and the related flow is the well known  Tzitzeica equation (TE)
\begin{equation}
v_{x\tau}=\mbox{\rm e}^v+\mbox{\rm e}^{-\frac{v}2}.   \label{ZE}
\end{equation}
It is obviously that the TE \eqref{ZE} is relativistically invariant (the space-time ($x$-$\tau$) exchange invariance under the light cone coordinate). Thus, similar to the sG equation, all symmetries of the Tzitzeica equation continue to be  symmetries under the $x\leftrightarrow\tau$ exchange transformation.

Based on the relativistic invariance of TE, we formulate the following proposition.
\begin{prop}  If $\sigma$ is a symmetry of TE, i.e., a solution of
\begin{equation}
\sigma_{x\tau}=\sigma\mbox{\rm e}^v-\frac{\sigma}2\mbox{\rm e}^{-\frac{v}2},\label{ZES}
\end{equation}
then
\begin{equation}
\Psi_D\Psi_{\Delta}\sigma=\sigma=\Psi_{\Delta}\Psi_D\sigma,\label{ZEPsi}
\end{equation}
where
\begin{equation}
\Psi_{\Delta}=\Delta_gg^{-1}\Delta_{g^3}^{-1}\Delta_g^5g^{-2}\Delta_{g^2}^{-1}\Delta_g^2,  \ \Delta_f\equiv f\partial_{\tau},\ \Delta_{f}^{-1}\equiv (\Delta_f)^{-1}=\partial_{\tau}^{-1}f^{-1}.   \label{PsiNa}
\end{equation}\end{prop}

A similar argument as we made for proving the proposition 2 may be employed to prove above proposition, namely taking the Tzitzeica equation \eqref{ZE} and the linearized symmetry equation \eqref{ZEPsi} into consideration  and integrations by parts. The detailed calculations, which are rather cumbersome, may be implemented with the assistant of the  computer algebras such MAPLE or MATHEMATICA. Thus, we omit the proof.

Thanks to the proposition 3, the dual hierarchy of the first positive hierarchy \eqref{PFG1} can be written as
\begin{equation}
v_{\tau_{6n+1}}=\Psi_{\Delta}^n v_{\tau}, \ n=0,\ 1,\ 2,\ \ldots . \label{PTE1}
\end{equation}

The first two of \eqref{PTE1} read as
\begin{equation*}
v_{\tau_{1}}=v_{\tau}=\partial_x^{-1}\left(\mbox{\rm e}^v+\mbox{\rm e}^{-\frac{v}2}\right), \label{TE1}
\end{equation*}
\begin{eqnarray*}
v_{\tau_{7}}&=&v_{\tau\tau\tau\tau\tau\tau\tau}-\frac74(v_{\tau}^2+2v_{\tau\tau})v_{\tau\tau\tau\tau\tau}
-7(v_{\tau}v_{\tau\tau}+v_{\tau\tau\tau})v_{\tau\tau\tau\tau}
\nonumber\\
&&-\frac78v_{\tau\tau\tau}(6v_{\tau}v_{\tau\tau\tau}-2v_{\tau}^2v_{\tau\tau}
-v_{\tau}^4+8v_{\tau\tau}^2)
+\frac74v_{\tau}^3v_{\tau\tau}^2+\frac76v_{\tau}v_{\tau\tau}^3-\frac1{48}v_{\tau}^7 \nonumber\\
&=&\Psi_D^{-1}\partial_x^{-1}\left(\mbox{\rm e}^v+\mbox{\rm e}^{-\frac{v}2}\right)=\Psi_{\Delta}v_{\tau}. \label{TE7}
\end{eqnarray*}
For the second positive hierarchy \eqref{PFG2}, its dual hierarchy is
\begin{equation*}
v_{\tau_{6n+5}}=\Psi_{\Delta}^{n} \left(v_{\tau\tau\tau\tau\tau}-\frac52v_{\tau\tau}v_{\tau\tau\tau}-\frac54v_{\tau}v_{\tau\tau}^2
-\frac54v_{\tau}^2v_{\tau\tau\tau}+\frac1{16}v_{\tau}^5\right),\ n=0,\ 1,\ 2,\ \ldots , \label{PTE2}
\end{equation*}
whose simplest model of the hierarchy is the fifth order negative PFG equation
\begin{eqnarray}
v_{\tau_{5}}&=&v_{\tau\tau\tau\tau\tau}-\frac52v_{\tau\tau}v_{\tau\tau\tau}-\frac54v_{\tau}v_{\tau\tau}^2
-\frac54v_{\tau}^2v_{\tau\tau\tau}+\frac1{16}v_{\tau}^5=K_5^{-}\label{NPFG}\\
&=&\partial_1^{-1}z^4+2\partial_2^{-1}\partial_1^{-1}z^3
+24\partial_1^{-1}\partial_2^{-3}w+12\partial_2^{-1}\partial_1^{-1}\partial_2^{-2}w \nonumber
\end{eqnarray}
with
$$ z_x=\mbox{\rm e}^{-\frac{v}2},\ w_x= \mbox{\rm e}^v.$$

\section{Duality of the dKPL hierarchy and  the two-dimensional dispersionless Toda hierarchy}

In (2+1)-dimensional cases, it is still possible to construct a positive integrable hierarchy with help of the FSSA \cite{FSSA} for certain nonlinear systems in the form
\begin{equation}
u_{tx_1}=K(t, x_1,\ldots,x_{m},u,u_{x_1},u_{x_2},\ldots)\equiv K(u),\ m\geq2, \label{2D}
\end{equation}
where $K(u)$ is a function of the space time ($\{t, x_1,\ldots,x_{m}\}$) and space ($\{ x_1,\ldots,x_{m}\}$) derivations of $u$ but not dependent on time ($t$) derivatives of $u$.
The (2+1)-dimensional nonlinear system \eqref{2D} possesses a formal series symmetry thanks to the following proposition.
\begin{prop}{\rm \cite{FSSA}} Let $f= f(t)$ and $g=g(t, x_2, \ldots,\ x_m)$ be arbitrary functions of the indicated variables, then
\begin{equation}
\sigma(f,g)=\sum_{k=0}^{\infty}f^{(-k)}(\partial_x^{-1}K'-\partial_t)^kg \label{sfg}
\end{equation}
is a formal symmetry of \eqref{2D} with $f^{(-k)}\equiv \partial_t^{-k}f$ and
$$K'h\equiv \lim_{\epsilon\rightarrow0} \frac{\rm d}{\rm d\epsilon}K(u+\epsilon h).$$\end{prop}

While it is not possible to prove the convergence for the formal series symmetry \eqref{sfg} with arbitrary $f$ and $g$ in general, it is fortunate and interesting that for various (2+1)-dimensional integrable systems such as the KP equation \cite{FSKP}, the Toda field \cite{FSSA}, the Nizhnik-Novikov-Veselov equation \cite{NNV}, the dispersive long wave equation \cite{DLWE} and the differential-difference Toda equation  \cite{DDToda}, the formal series symmetries can be truncated to a closed summation form by selecting the arbitrary function $g$ as special polynomial functions.

Whence the formal series symmetry \eqref{sfg} is truncated up to the $n$th term by fixing $g=g_n$, we can rewrite \eqref{sfg} as
\begin{equation}
\sigma_n(f)=\sum_{k=0}^{n}f^{(n-k)}(\partial_x^{-1}K'-\partial_t)^kg_n, \label{sfgn}
\end{equation}
after changing $f$ to $f^{(n)}$ because of its arbitrariness. More specifically, by fixing the arbitrary function $f$ as a constant, say, $f=1$, one may obtain an integrable (positive) hierarchy
\begin{equation}
u_{t_n}=(\partial_x^{-1}K'-\partial_t)^ng_n \label{ph}
\end{equation}
if \eqref{2D} is integrable.

Now we consider the following equation
\begin{equation}
v_{yt}=(vv_{x})_y-v_{xx},  \label{dKPv}
\end{equation}
which was proposed recently by Zakharov  et al. \cite{dKPL}.  This (2+1)-dimensional equation, referred as the dispersionless Kadomtsev-Petviashvili-like (dKPL) equation,  resembles
the dispersionless Kadomtsev-Petviashvili (dKP) equation
\begin{equation}
v_{xt}=(vv_{x})_x-v_{yy}, \label{dKPv1}
\end{equation}
and the dispersionless negative BKP equation
\begin{equation}
P_{yt}=3(P_xP_{y})_x-3P_{xx}. \label{dKPv2}
\end{equation}
It is noted that  \eqref{dKPv}, \eqref{dKPv1} and \eqref{dKPv2} are different (2+1)-dimensional extensions of the Riemann equation  $w_t=ww_x$.

The potential form of the dKPL equation \eqref{dKPv} reads as
\begin{equation}
u_{yt}=\frac12u_{xy}^2-u_{xx}\equiv G, \label{dKP}
\end{equation}
where $v=u_{xy}$. Applying the proposition 4 to it
and fixing the arbitrary function $g$ as $g=-\frac1{4n!}x^n$, we find an integrable positive potential dKPL hierarchy in the form
\begin{equation}
u_{t_{n-2}}=-\frac{1}{4n!}(\partial_y^{-1}u_{xy}\partial_{x}\partial_{y} -\partial_y^{-1}\partial_x^2-\partial_t)^{n-1}x^n,\ n=3,\ 4,\ \cdots . \label{dKPh}
\end{equation}

The first four flows of this hierarchy \eqref{dKPh} (for $n=3,\ 4,\ 5$ and $6$, respectively) read
\begin{equation*}
u_{t_1}=\frac14u_x,\ \label{dKP1}
\end{equation*}
\begin{equation*}
u_{yt_2}=\frac12u_{xy}^2-u_{xx},\ \label{dKP2}
\end{equation*}
which is just the potential dKPL equation \eqref{dKP} with $t_2=t$,
\begin{equation}
u_{yt_3}=\frac16u_{xy}^3-\frac12u_{xx}u_{xy}-\frac12v,\ v_y=G_x,\ \label{dKP3}
\end{equation}
with $G$ being defined in \eqref{dKP} and
\begin{equation*}
u_{yyt_4}=\frac34(u_{xyy}-\partial_x)(v+2u_{xx}u_{xy}-u_{xy}^3).\label{dKP4}
\end{equation*}

In general, recursion operators are not available for (2+1)-dimensional integrable systems, so we have to adopt other methods \cite{NH1,BT1NH2}  to find the related
 negative hierarchy. For a given (2+1)-dimensional system, provided that a symmetry flow,
$$u_{x\tau}=F(u),$$
which possesses the space-time $\{x,\tau\}$ exchange invariance,  then we may construct a dual negative hierarchy by using the duality relation $u_{\tau}=\partial_x^{-1}F(u)$. Fortunately, for the potential dKPL equation \eqref{dKP} we have the following proposition.
\begin{prop}

The equation
\begin{equation}
u_{\tau}=\partial_x^{-1}{\rm e}^{-u_{yy}} \label{TodaDual}
\end{equation}
constitutes  a symmetry of the potential dKPL equation \eqref{dKP}.%, i.e., a solution of

\end{prop}
Proof. To prove it, we need to show $\sigma=\partial_x^{-1}{\rm e}^{-u_{yy}}$ solves the linearized potential dKPL equation, namely
\begin{equation}
\sigma_{yt}-u_{xy}\sigma_{xy}+\sigma_{xx}=0. \label{dKPsigma}
\end{equation}
%the proposition 5, one can use some alternative ways, say, (i) check $u_{ytx\tau}=u_{x\tau yt}$, (ii) check $u_{t\tau}=u_{\tau t}$ and (iii) check \eqref{dKPsigma} with $\sigma=\partial_x^{-1}{\rm e}^{-u_{yy}}$. Here, we use the third way to prove the proposition.
Substituting $\sigma=\partial_x^{-1}{\rm e}^{-u_{yy}}$ into the left-hand side of \eqref{dKPsigma}, we have
\begin{eqnarray}
(u_{xy}u_{yyy}-u_{xyy}){\rm e}^{-u_{yy}}+\partial_x^{-1}\left[(u_{yyy}u_{yyt}-u_{yyyt}){\rm e}^{-u_{yy}}\right]. \label{pr1}
\end{eqnarray}
After using the potential dKPL equation, \eqref{pr1} becomes
\begin{eqnarray*}
\lefteqn{(u_{xy}u_{yyy}-u_{xyy}){\rm e}^{-u_{yy}} +\partial_x^{-1}\left[{\rm e}^{-u_{yy}}(u_{yyy}\partial_y-\partial_{y}^2)G\right]}\nonumber\\
&&=\partial_x^{-1}\left\{\partial_x\left[(u_{xy}u_{yyy}-u_{xyy}){\rm e}^{-u_{yy}}\right] +{\rm e}^{-u_{yy}}(u_{yyy}\partial_y-\partial_{y}^2)G\right\}\nonumber\\
&&=\partial_x^{-1}\left[(u_{yyy}-\partial_y)(G_y+u_{xxy}-u_{xy}u_{xyy}){\rm e}^{-u_{yy}}\right]
\label{pr2},
\end{eqnarray*}
which vanishes due to the definition of $G$. Thus the proposition 5 is proved.

The equation \eqref{TodaDual} or
\begin{equation}
u_{x\tau}={\rm e}^{-u_{yy}}, \label{Toda}
\end{equation}
is  the large $N$ limit ($N\rightarrow \infty$) of the two-dimensional  sl$(N+1$) Toda field \cite{Toda,To2,To3,ma} which is also known as  the Boyer-Finley equation \cite{BF} or the SU$(\infty$) Toda equation \cite{Ward} or the two-dimensional dispersionless Toda (2ddT) equation \cite{To4} . Because of the relativistic invariance of the 2ddT equation \eqref{Toda}, we can build a  hierarchy (negative dKPL hierarchy) which is a dual hierarchy of the positive potential dKPL hierarchy \eqref{dKPh}
 \begin{equation}
u_{\tau_{n-2}}=-\frac{1}{4n!}(\partial_y^{-1}u_{y\tau}\partial_{\tau}\partial_{y} -\partial_y^{-1}\partial_\tau^2-\partial_t)^{n-1}\tau^n,\ n=3,\ 4,\ \cdots, \label{NdKPh}
\end{equation}
with $\tau$ being defined by the duality relation \eqref{TodaDual}.

The first three equations of the hierarchy \eqref{NdKPh} possess the following forms
\begin{equation*}
u_{x\tau_1}=\frac14{\rm e}^{-u_{yy}}=\frac14u_{x\tau},\ \label{NdKP1}
\end{equation*}
which is just the 2ddT equation \eqref{Toda} with $\tau_1=4\tau$,
\begin{eqnarray*}\label{NdKP2}
u_{xy\tau_2}=(zu_{yyy}-z_y){\rm e}^{-u_{yy}},\ z_x=u_{yyy}{\rm e}^{-u_{yy}},
\end{eqnarray*}
and
\begin{eqnarray*}\label{NdKP3}
u_{xyy\tau_3}=z_xw_y+\frac12wz_{xy}-\frac12w_{yy}{\rm e}^{-u_{yy}},\ w_x=(z_y-2zu_{yyy}){\rm e}^{-u_{yy}}.
\end{eqnarray*}

\section{Duality hierarchies from the two-dimensional Toda lattice}

It is known that the 2ddT equation \eqref{Toda} is an integrable continuous limit of the following differential-difference system \cite{DdToda,DDToda1,DToda,dToda}
\begin{eqnarray}\label{DDToda}
u_{x\tau}={\rm e}^{u_{n-1}-u_{n}}-{\rm e}^{u_{n}-u_{n+1}}\equiv A-A_1=-\Delta A_1,\ A_k\equiv {\rm e}^{-\Delta u_{n+k}},\ \Delta f_n=f_n-f_{n-1},
\end{eqnarray}
which is the celebrated two-dimensional Toda lattice (2dTL).  In this section, we aim to construct possible dual systems by  taking the 2dTL \eqref{DDToda}, which is relativistically invariant,  as a duality relation.
% Now, the question is can we get something by using \eqref{DDToda} as a duality relation.
To this end, we should first build a  hierarchy (negative hierarchy) related to \eqref{DDToda} and this will be done by means of  the mastersymmetry method \cite{Master}.

% Then we can find the related dual positive hierarchy by using the relativistic invariant Toda equation \eqref{DDToda} as the duality relation.
%Here, we use the mastersymmetry method \cite{Master} to find next higher order Toda lattice system.

A direct calculation yields
\begin{eqnarray}\label{Mastx}
[n,K]=K_1=K=\partial_x^{-1}\left[{\rm e}^{u_{n-1}-u_{n}}-{\rm e}^{u_{n}-u_{n+1}}\right]=u_\tau=u_{\tau_1},
\end{eqnarray}
where the commutator is defined as
\begin{eqnarray*}\label{Commun}
[F,G]=F'G-G'F=\lim_{\epsilon=0}\frac{\rm d}{\rm d\epsilon}\left[F(u_n+\epsilon G)-G(u_n+\epsilon F)\right].
\end{eqnarray*}
Therefore, while $n$ is not a symmetry of the 2dTL \eqref{DDToda}, it is actually a mastersymmetry.
To find the next (higher order) flow one can apply the higher order mastersymmetry $n^2$. Some simple calculations result in
\begin{eqnarray}\label{Mast2}
[[n^2,K],K]=2\Delta\partial_x^{-1}A_1\partial_x^{-1}(A_{2}-A)\equiv K_2=u_{\tau_2},
\end{eqnarray}
which, after taking the 2dTL  \eqref{DDToda} into consideration, may be rewritten as
\begin{eqnarray}\label{Toda2}
u_{x\tau_2}=2(u_{n}+u_{n-1})_{\tau}{\rm e}^{u_{n-1}-u_{n}}-2(u_{n}+u_{n+1})_\tau {\rm e}^{u_n-u_{n+1}}=-4\Delta A_1 E u_{n+1,\tau},\
\end{eqnarray}
where the average operator $E$ is defined by
$$Ef_n=\frac12(f_n+f_{n-1})$$
and $\tau$ is related to $x$ by the 2dTL \eqref{DDToda}. Under  $x=\tau$, \eqref{Toda2} reduces to an equation  appeared early \cite{Cao1,HuXB}.

Now taking \eqref{DDToda} as a duality relation, we may work out the dual equation (the equation of the related positive hierarchy) of \eqref{Toda2} in the $\{\tau,\ \tau_2\}$ space. Indeed, eliminating $u_x$ by means of \eqref{DDToda}, \eqref{Toda2} becomes
\begin{eqnarray}\label{DdKPL}
\Delta \left\{A_1\left[\Delta (u_{n+1,\tau_2}-2u_{n+1,\tau}^2)+4Eu_{n+1,\tau\tau}\right]\right\}=0,
\end{eqnarray}
which leads to %\eqref{DdKPL} is equivalent to
\begin{eqnarray*}\label{DdKP}
\Delta u_{n\tau_2 }=2 \Delta u_{n\tau }^2-4E u_{n\tau\tau}
\end{eqnarray*}
or
\begin{eqnarray*}\label{DdKPa}
u_{n\tau_2 }=2 u_{n\tau }^2-4\Delta^{-1} E u_{n\tau\tau},\ \Delta^{-1}\Delta=1.
\end{eqnarray*}
Applying the relativistic invariance again, we have another but equivalent integrable model
\begin{eqnarray}\label{DdKPb}
u_{n\tau_2 }=2 u_{nx }^2-4\Delta^{-1} E u_{nxx} \equiv K_{2}^+.
\end{eqnarray}
It is interesting to note that  the new equation \eqref{DdKPb} just constructed may be taken as  a semi-discrete  potential dKPL equation \eqref{dKP} where the variable $y$ is discretized.

Directly we may check that
\begin{eqnarray}\label{KK'}
u_{n\tau_2\tau}-u_{n\tau\tau_2}=[K_2^+,K]=0
\end{eqnarray}
holds, which also implies $[[[n^2,K],K],K]=0$. In other words, both $K_2$ and $K_2^+$ defined in \eqref{Mast2} and \eqref{DdKPb} are symmetries and $n^2$ is a mastersymmetry of the 2dTL \eqref{DDToda}.

\section{From the real second heavenly equation to dual hierarchies}

In this section, we apply the similar duality approach to find positive and negative heavenly hierarchies by means of the exchange invariance
\begin{equation}\label{ExCh}
\{x,\ \tau,\ y,\ z,\ u\}\longleftrightarrow \{\tau,\ x,\ z,\ y,\ -u\}
\end{equation}
of the real second heavenly equation \cite{Heaven1}
\begin{equation}\label{heaven}
u_{xz}-u_{y\tau}+u_{zz}u_{yy}-u_{yz}^2=0.
\end{equation}
The heavenly equations, introduced in \cite{Heaven1} by Plebanski, describe self-dual vacuum  solutions  of  the  Einstein  equations. The equation \eqref{heaven} has been studied extensively and many results have been established (see \cite{Heaven,Heaven2,Heaven3} and the references therein). In particular, the multi-Hamiltonian structures and assocaited (positive) hierarchies have been explored by different methods \cite{Heaven3,Heaven4, Heaven5}.
%have beenThe positive hierarchies of the heavenly equation \eqref{heaven} have been studied by several authors because of the existence of the bi-Hamitonian structure and/or the formal recursion operators\cite{Heaven3}.

According to the FSSA and the mastersymmetry method as discussed in the sections IV and V, we can find that the heavenly equation possesses two positive hierarchies in the forms
\begin{eqnarray}\label{sf}
u_{t_{n-1}}=\left[\partial_y^{-1}(\partial_x\partial_z+u_{zz}\partial_y^2 +u_{yy}\partial_z^2-2u_{yz}\partial_y\partial_z)-\partial_{\tau}\right]^{n-1}\frac{z^n}{n!}, \ n=1,\ 2,\ \ldots,
\end{eqnarray}
and
\begin{eqnarray}\label{sF}
u_{\tau_{n-1}}=\left[\partial_z^{-1}(\partial_\tau \partial_y-u_{zz}\partial_y^2 -u_{yy}\partial_z^2+2u_{yz}\partial_y\partial_z)-\partial_x\right]^{n-1}\frac{y^n}{n!},  \ n=1,\ 2,\ \ldots.
\end{eqnarray}
It is clear that above two hierarchies are related via the discrete symmetry transformation \eqref{ExCh}.

In the positive hierarchy \eqref{sf} the variable $\tau$ should be eliminated by means of the heavenly equation \eqref{heaven}, i.e.,
\begin{eqnarray}\label{HEtau}
u_{\tau}=\partial_y^{-1}(u_{xz}+u_{zz}u_{yy}-u_{yz}^2)\equiv K.
\end{eqnarray}
For the second positive hierarchy \eqref{sF}, we should eliminate the variable $x$ via
\begin{eqnarray}\label{HEx}
u_{x}=\partial_z^{-1}(u_{y\tau}-u_{zz}u_{yy}+u_{yz}^2)\equiv P.
\end{eqnarray}

After some tedious calculations, we find that the hierarchy \eqref{sf} may be reformulated concisely as
\begin{eqnarray}\label{utn}
&&u_{t_n}=\Phi^{n} z=\Phi^{n-2} u_x,\ \Phi\equiv \partial_y^{-1}(u_{yy}\partial_z-u_{yz}\partial_y+\partial_x)
\end{eqnarray}
which for $n=1, 2, 3$ leads to
\begin{eqnarray}
&&u_{t_1}=u_y=\Phi z,\ \nonumber\\
&&u_{t_2}=u_x=\Phi^2 z,\nonumber \\ %\label{ut2}
&&u_{yt_3}=u_{yy}u_{xz}-u_{yz}u_{xy}+u_{xx}= (\Phi^3 z)_y,\ \label{ut3}
\end{eqnarray}
while the second hierarchy \eqref{sF} can be written as
\begin{eqnarray}\label{uTn}
&&u_{\tau_n}=\Psi^{n} y=\Psi^{n-2} u_z,\ \Psi\equiv \partial_z^{-1}(\partial_{\tau}-u_{zz}\partial_y+u_{yz}\partial_z)
\end{eqnarray}
with the first three examples
\begin{eqnarray*}
&&u_{\tau_1}=-u_z=\Psi y,\ \nonumber\\
&&u_{\tau_2}=-u_{\tau}=\Psi^2 y, \label{uT2}\\
&&u_{z\tau_3}=u_{yy}u_{z\tau}-u_{yz}u_{y\tau}-u_{\tau\tau}= (\Psi^3 y)_z.\ \label{uT3}
\end{eqnarray*}

The dual negative hierarchy of \eqref{utn} can be obtained by using the relativistic invariance \eqref{ExCh}. The result reads as
\begin{eqnarray}\label{nutn}
&&u_{\tau_n}=\left.\Psi^{n-2} u_z\right|_{u_{\tau}=K},
\end{eqnarray}
and explicitly its first nontrivial flow (for $n=3$) is
\begin{eqnarray*}
u_{y\tau_3}&=&\partial_z^{-1}
\left.(u_{y\tau}u_{zz}-u_{yz}u_{z\tau}-u_{\tau\tau})_y\right|_{u_{\tau}=K}\nonumber\\
&=&v_zu_{yy}-v_yu_{zy}+v_x, \label{utau3}\\
v_y&=&u_{yz}^2-u_{yy}u_{zz}-u_{zx}.
\end{eqnarray*}

In the same way, the dual negative hierarchy of  \eqref{uTn} has the form
\begin{eqnarray}\label{nuTn}
&&u_{t_n}=\left.\Phi^{n-2} u_x\right|_{u_{x}=P}.
\end{eqnarray}

Above discussions indicate that both $\Phi$ and $\Psi$ should be  the recursion operators of the second heavenly equation \eqref{heaven}. Indeed, we have
\begin{prop}
 Let $\sigma$ be a symmetry of the second heavenly equation \eqref{heaven}, i.e., a solution of
\begin{equation}\label{HEsym}
 \sigma_{y\tau}-\sigma_{xz}-\sigma_{zz}u_{yy}-u_{zz}\sigma_{yy} +2u_{yz}\sigma_{yz}=0,
\end{equation}
so are $\Phi\sigma$ and $\Psi\sigma$. \end{prop}
Proof. Making the change $\sigma\rightarrow \Phi \sigma$ and eliminating $\sigma_{\tau}$ and $u_{\tau}$ by means of \eqref{HEsym} and \eqref{heaven}, we have
\begin{eqnarray}\label{Phisym}
\lefteqn{\left.\partial_y(\partial_{\tau}-K')\Phi\sigma\right|_{\eqref{heaven}\eqref{HEsym}}}\nonumber\\
&&=u_{yy}\left\{(u_{yz}\sigma_z-u_{zz}\sigma_y)_z  +\partial_y^{-1}\left[(u_{zz}\sigma_{yy}-u_{yyz}\sigma_{z})_z -u_{yz}\sigma_{yzz} +u_{yzzz}\sigma_y\right]\right\}\nonumber\\
&&=u_{yy}\partial_y^{-1}\left[\partial_y(u_{yz}\sigma_z-u_{zz}\sigma_y)_z  +(u_{zz}\sigma_{yy}-u_{yyz}\sigma_{z})_z -u_{yz}\sigma_{yzz} +u_{yzzz}\sigma_y\right]\nonumber\\
&&=u_{yy}\partial_y^{-1} 0 =0.
\end{eqnarray}
So $\Phi \sigma$ is also a symmetry of \eqref{heaven}. The conclusion for $\Psi \sigma$  also holds due to \eqref{ExCh}. Thus the proposition is proved.

 Several remarks are in order:

\bf Remark 1. \rm Though the hierarchy \eqref{nutn} is a dual hierarchy of \eqref{utn}, we should mention that $\Psi$ is not a inverse of $\Phi$.

\bf Remark 2. \rm While we constructed the operators $\Phi$ and $\Psi$ via the FSSA, they did appear in a early work  by Dunajski and Mason \cite{Heaven3, Heaven4}.

\bf Remak 3. \rm Two recursion operators $\Phi$ and $\Psi$ commute each other, i.e.,
$$[\Phi,\ \Psi]\sigma=(\Phi\Psi-\Psi\Phi)\sigma=0, $$
where $\sigma$ is a symmetry of the second heavenly equation \eqref{heaven}.

According to the remark 3, two hierarchies \eqref{utn} and \eqref{uTn} can be uniformly written as
\begin{eqnarray*}\label{utTn}
&&u_{t_{nm}}=\left. \Phi^{n}\Psi^{m} \sigma_{0,0}\right|_{\eqref{heaven}},\ \sigma_{0,0}= x\ {\rm or } \ \tau.
\end{eqnarray*}
while the dual hierarchy of \eqref{utTn} reads
\begin{eqnarray}\label{uTtn}
&&u_{t_{mn}}=\left. \Phi^{m}\Psi^{n} \sigma_{0,0}\right|_{\eqref{heaven}},\ \sigma_{0,0}= \tau\ {\rm or } \ x.
\end{eqnarray}

\section{Summary and discussions}
In summary, if there is a relativistically invariant flow for an integrable nonlinear system, then the related positive and negative hierarchies are dual each other simply by taking the relativistically invariant flow as the duality relation. In (1+1)-dimensional cases, the positive hierarchies are local in a proper space ($\{x,\ t_n\}$) and nonlocal in its dual space ($\{\tau,\ t_n\}$) where the negative hierarchies are local in the space $\{\tau,\ \tau_n\}$ and nonlocal in the dual space $\{x,\ \tau_n\}$. Because of the existence of the recursion operators for (1+1)-dimensional integrable systems, we find that the recursion operators and their inverses and then the positive hierarchies and negative hierarchies possess completely same forms but with different ``space" variables which are linked each other by means of the relativistic duality relations. This special structure will undoubtedly bring a lot of convenience when we deduce the integrable properties of the whole hierarchies under consideration. For example, by the dependent variable transformation, we can establish the following unified bilinear form for the whole potential mKdV hierarchy
\begin{eqnarray*}
&&(D_{2n+1}-D_x^2D_{2n-1})f^*\cdot f=0,\\
&&(D_{-(2n+1)}-D_\tau^2D_{-(2n-1)})f^*\cdot f=0,\\
&&D_x^2f^*\cdot f=0, \;\;
D_\tau^2f^*\cdot f=0\\
&&D_xD_\tau f\cdot f=\frac 12(f^2-{f^*}^2)
\end{eqnarray*}
where $D_{2k+1}\equiv D_{t_{2k+1}}$ and $D_{-(2k+1)}\equiv D_{\tau_{2k+1}}$ are Hirota's bilinear operators defined by
\begin{eqnarray*}
&& D_t^mD_x^n a(t,x)\cdot b(t,x)=\frac{\partial^m}{\partial s^m}\frac{\partial^n}{\partial y^n}a(t+s,x+y)b(t-s,x-y)|_{s=0,y=0},\nonumber\\
&& \qquad \qquad \qquad \qquad \qquad \qquad \qquad \qquad \qquad
m,n=0,1,2,\cdots.
\end{eqnarray*}
Furthermore, starting from unified  bilinear form for the whole potential mKdV hierarchy, we can derive the B\"acklund transformation and nonlinear superposition formula.
 Besides, it might be of interest to study the following equations of $x\leftrightarrow y$ invariance from geometric point of view,
\begin{eqnarray*}
v_{t}&=&v_{xxx}+v_{yyy}+\frac12\big({v_x}^3+{v_{y}}^3\big),\ \ v_{xy}=\sin v,\\
v_{z t}&=&\big(v_{xz}^2+v_{yz}^2\big)-2(v_{xx}+v_{yy}),\ \ v_{xy}={\rm e}^{-v_{zz}},\\
 v_{nt} &=&\big(v_{ny }^2+v_{nx}^2\big)-2\Delta^{-1} E \big(v_{nyy}+v_{nxx}\big),\
 v_{n x y}={\rm e}^{v_{n-1}-v_n}-{\rm e}^{v_{n}-v_{n+1}},\\
 v_{t}&=&v_{xxxxx}-\frac54\big(2v_{xx}v_{xxx}+v_xv_{xx}^2+v_x^2v_{xxx}\big)+\frac1{16}v_x^5  \\
&&
+v_{yyyyy}-\frac54\big(2v_{yy}v_{yyy}+v_{y}v_{yy}^2 +v_{y}^2v_{yyy}\big)+\frac1{16}v_{y}^5,\  v_{xy}={\rm e}^v+{\rm e}^{-v/2}.
\end{eqnarray*}

For a (2+1)-dimensional integrable model, the formal series symmetry approach \cite{FSSA} and the mastersymmetry method \cite{Master} can be readily used to find positive hierarchies. The negative hierarchies can be obtained by means of  Lax operators \cite{NH1} or the nonlocal symmetries like the squared eigenfunction symmetries and infinitesimal B\"acklund/Darboux transformations \cite{NH2}. In this paper, the positive dKPL hierarchy is constructed  within the framework of the FSSA while the negative hierarchy %(the (2+1)-dimensional continuous Toda hierarchy)
is found by the duality relation owing to the relativistic invariance of the two-dimensional dispersionless Toda equation. On the other hand, if one has a relativistically invariant integrable system, then it is possible to directly find the related dual hierarchies by using the system as duality relation. By combining the mastersymmetry method and the duality approach, the differential-difference dKPL equation \eqref{DdKPL} and the related higher order dual Toda equation are successfully obtained from the two-dimensional Toda lattice. In fact, there are many other relativistically invariant integrable systems such as the coupled Tzitzeica-sinh-Gordon model \cite{FG80}, the multicomponent sinh-Gordon systems \cite{FG80}, the Pohlmeyer-Lund-Regge-Getmanov model \cite{CsG,CsG1,GsG2}, the principal SU(n) chiral model \cite{Chiral}, the massive Thiring model  \cite{Thiring}, O(n) nonlinear $\sigma$ model \cite{CsG}, the self-dual Yang-Mills equation \cite{Ward81} and so on. For those relativistically invariant integrable systems, it is interesting to construct the related positive and negative dual hierarchies by adopting the duality method proposed in this paper.

In higher dimensions, the second heavenly equation \eqref{heaven} is especially important because it can be derived from both the Einstein's general relativistic equation and the self-dual Yang-Mills equation. Though the recursion operators of the second heavenly equation have been studied by Dunajski and Mason  via the twistor theory of the  anti-self-dual Einstein vacuum equations \cite{Heaven3, Heaven4}, those operators %two much simpler forms of the recursion operators of the second heavenly equation \eqref{heaven}
are reobtained here directly by the simple FSSA. Using the heavenly equation as a duality relation the positive and negative second heavenly hierarchies are naturally obtained. Especially, beyond the known classifications  \cite{Novikov}, a further simple second order integrable system \eqref{ut3}  appears in the positive heavenly hierarchy \eqref{utn} or \eqref{uTtn}.

\section*{Acknowledgements}
%The author is grateful to thank Professors X. Y. Tang, D. J. Zhang, Z. N. Zhu, Q. P. Liu, X. B. Hu, Y. Q. Li and Y. Chen for their helpful discussions.
The work was sponsored by the National Natural Science Foundations of China (Nos. 11975131, 11871471  and  11931017) and K. C. Wong Magna Fund in Ningbo University.

\end{document}